\documentclass[conference]{IEEEtran} 
\usepackage{amsmath}
\usepackage{amssymb}
\usepackage{tabularx}
\usepackage{booktabs}
\usepackage{longtable}
\usepackage{mathtools}
\usepackage{graphicx}
\usepackage{optidef}
\usepackage{amsfonts}
\usepackage{algorithm}
\usepackage{algorithmic}
\usepackage{caption}
\usepackage{svg}
\usepackage{subcaption}
\usepackage{amsthm}

\usepackage{xcolor}
\usepackage{breqn}
\usepackage{cuted}
\usepackage{balance}
\usepackage{float}
\usepackage{listings}
\usepackage{capt-of}

\usepackage[top=0.75in, left=0.64in, right=0.64in, bottom=1.04in, letterpaper]{geometry}

\ifCLASSINFOpdf
\else
\fi
\DeclareUnicodeCharacter{2212}{-}

\begin{document}

%
\title{Handling Interoperability Issues in 6G ORAN: Lessons Learned from Research Lab-based Integrations}
%
%
%
\author{Poonam Yadav, \IEEEauthorblockN {Yifan Liu, Mohit Bidikar, Rana M. Sohaib, Yi Chu, David Grace}
\IEEEauthorblockA{University of York, York, UK\\ Emails:\{poonam.yadav, yifan.liu, mohit.bidikar, rana.sohaib, yi.chu, david.grace\}@york.ac.uk}}



\maketitle

\begin{abstract}
As 6G networks continue to evolve, the Open Radio Access Network (ORAN) framework is gaining prominence for its promise of network flexibility, vendor-neutral architecture, and enhanced service delivery. However, achieving true interoperability across diverse hardware and software components from multiple vendors remains a critical challenge. This research focuses on identifying and addressing interoperability issues within 6G ORAN environments, leveraging insights from research lab-based integrations. By systematically analysing lab-based case studies, the study aims to uncover the root causes of integration failures, propose solutions to ensure smooth multi-vendor compatibility, and establish a set of best practices for future deployments. This work will contribute to developing robust guidelines for 6G ORAN ecosystems, accelerating seamless and scalable deployments in real-world networks.
\end{abstract}


\IEEEpeerreviewmaketitle

\section{Introduction}
In recent years, the Open Radio Access Network (ORAN or O-RAN)  initiative \cite{oran_overview} has emerged as a critical force in the telecommunications industry to create more open, flexible, and interoperable Radio Access Networks (RANs). Traditionally, RANs have been tightly integrated systems, with hardware and software provided by a few major vendors, limiting flexibility and innovation. ORAN is revolutionizing this by decoupling hardware and software, enabling a mix of vendors and fostering more significant innovation in network infrastructure. This approach is significantly accelerating the deployment of 5G and beyond. 

Despite the maturity of ORAN specifications and increasing adoption in both research and industry, achieving true interoperability across components from different vendors remains a critical challenge. Issues such as inconsistent support for E2 interface, variations in implementation semantics, and closed software loops frequently arise during integration. These challenges are particularly pronounced in experimental or pre-commercial environments, where interoperability testing and debugging resources may be limited.

While previous work has addressed high-level architecture, marketplace ecosystems, and testbed design, fewer studies have deeply analyzed real-world integration efforts, especially those involving a mix of commercial and open-source platforms. This paper addresses that gap by reporting on hands-on integration experiences between:
\begin{itemize}
\item[(a)] A commercial E2 node and an open-source RIC;
\item[(b)] An open-source E2 node and an open-source RIC;
\item[(c)] A commercial E2 node and a commercial RIC;
\end{itemize}

Our work identifies failure points, interprets log-level behavior, and evaluates the effects of mismatched capabilities, versioning issues, and tool limitations. Beyond documenting the integration process, we propose actionable best practices for improving future ORAN lab deployments.\\


Our Contributions:

\begin{itemize}

\item[-] Unlike prior studies that focus predominantly on theoretical modeling, this work provides a hands-on perspective derived from extensive laboratory integrations involving open-source and commercial solutions.

\item[-] We debugged and documented the integration of RAN and RIC components from both open-source and commercial vendors.


\item[-] Throughout the integration process, we identify several challenges related to interoperability, resource allocation, latency, and performance optimisation. We offer solutions and recommendations to address these challenges, contributing to the ongoing development of O-RAN standards and best practices.

\item[-]  Our findings and proposed solutions help refine the O-RAN architecture, improving the overall design and operation of the near-RT RIC and xApps. This work supports the continued advancement of open, flexible, and efficient RAN systems, ultimately facilitating the deployment of next-generation 5G and beyond networks.

\end{itemize}

This paper is organized as follows: in Section \ref{sec:Background}, we briefly background and related work on advances in ORAN. In Section \ref{sec:Experiment}, we introduce the experimental setups including various components of ORAN for interoperability testing. Key challenges for interoperability are identified in Section \ref{sec:Findings}. Followed by potential future directions in Section \ref{sec:Future}.

\section{Background and Related Work} \label{sec:Background}
The Open Radio Access Network (ORAN) paradigm marks a significant shift toward disaggregated, intelligent, and vendor-neutral RAN architectures. As the telecommunications industry advances from 5G toward 6G, ORAN is emerging as a foundational enabler of flexibility, automation, and innovation in radio access networks. By standardizing interfaces between decoupled RAN components, ORAN allows operators to deploy interoperable solutions from multiple vendors, thereby fostering openness and ecosystem diversity.

In contrast to traditional RAN implementations, where hardware and software are tightly integrated and typically sourced from a single vendor, ORAN adopts a modular architecture composed of distinct, interoperable units. These include the Radio Unit (RU), Distributed Unit (DU), Centralized Unit (CU), and the Service Management and Orchestration (SMO) framework. Central to this architecture is the RAN Intelligent Controller (RIC), which is divided into two functional domains: the Non-Real-Time RIC (Non-RT RIC), which oversees high-level policy enforcement and AI/ML model lifecycle management, and the Near-Real-Time RIC (Near-RT RIC), which enables low-latency, closed-loop control of RAN behavior.

The Near-RT RIC achieves real-time optimization by executing microservices known as xApps. These applications monitor RAN state and issue control commands via the E2 interface. Internally, the Near-RT RIC contains key modules such as the E2 Termination (E2T), Subscription Manager, and Control Handler. These modules collectively manage session state, parse and validate protocol messages, and coordinate control signals dispatched to E2 nodes.

The E2 interface forms the core linkage between the Near-RT RIC and E2 nodes, which typically reside within the CU or DU. It supports various service models, including Key Performance Measurement (KPM), Radio Configuration (RC), and custom E2 Service Models (E2SM), allowing xApps to both consume RAN telemetry and issue control instructions. While this architecture theoretically enables dynamic and intelligent RAN behaviors, our experimental work reveals persistent interoperability issues. These include message encoding errors, mismatched expectations between vendors, and inconsistent support for service model extensions.
These findings motivate the need for deeper investigation into interface-level behaviors, which is the central focus of the present paper.
Similar efforts have been undertaken by the research community to implement and integrate ORAN components.

\begin{figure}[htbp]
\centering
\includegraphics[width=\columnwidth]{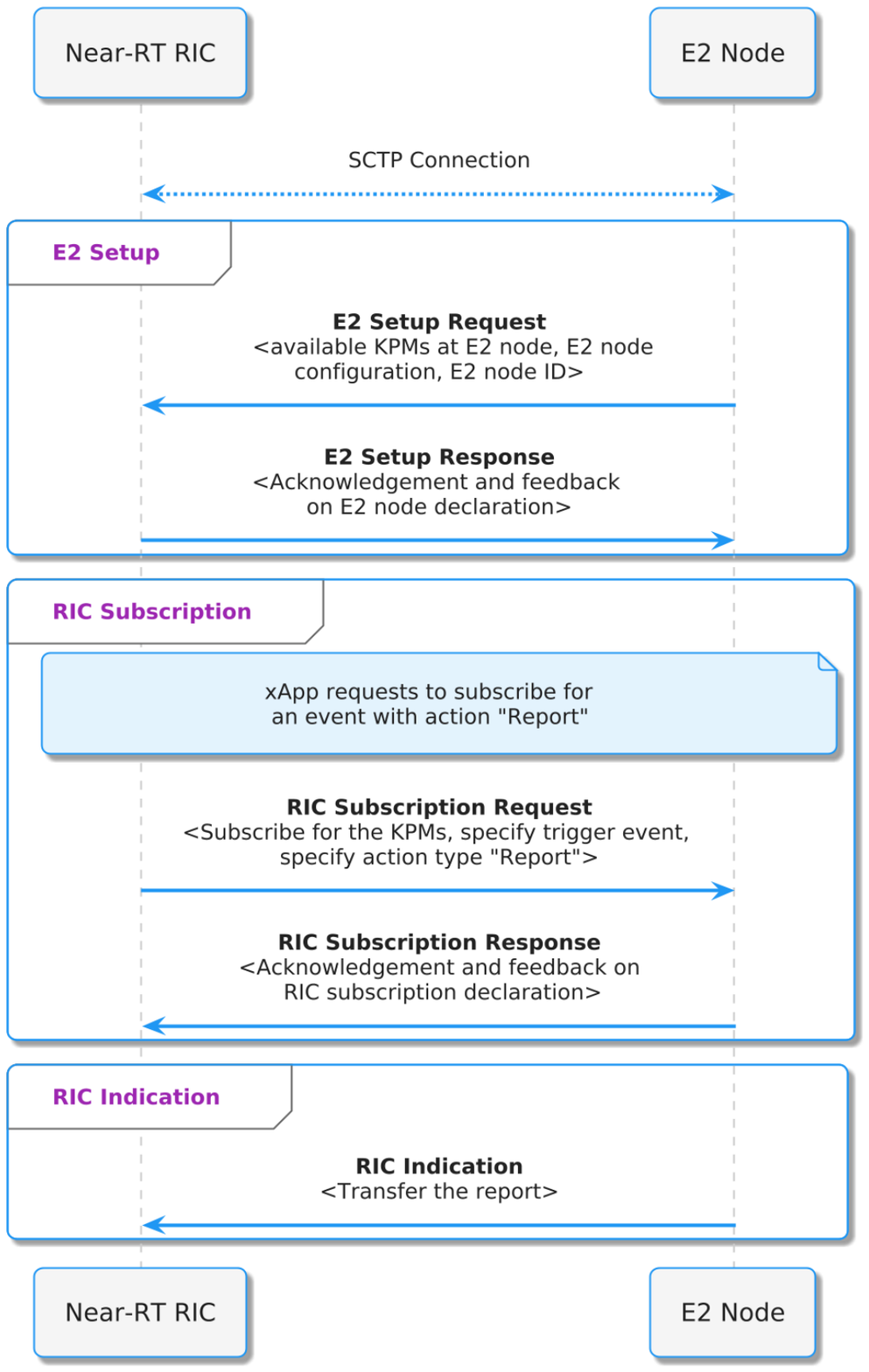}
\caption{E2 node interaction workflow with Near-RT RIC.}
\label{E2node} 
\end{figure}



Mehran et al.~\cite{mehran2024experimental} provide a comprehensive experimental evaluation of multi-vendor 5G ORAN deployments, identifying protocol-level inconsistencies and control plane mismatches as key challenges. 

Similarly, H\"oschele et al.~\cite{hoschele20225g} examine interoperability within large-scale ORAN testbeds, exposing frequent issues in E2 setup and service model alignment.

To address scale and flexibility, Kord et al.~\cite{kord2024evolving} propose a broad interoperability definition for Open RAN ecosystems, emphasizing the need for robust testing environments. Complementary work by Bonati et al.~\cite{bonati20245g} introduces 5G-CT, a test automation framework for end-to-end ORAN deployments, though it focuses primarily on deployment pipelines rather than low-level integration issues.

Several testbeds have been proposed to support ORAN experimentation. Villa et al.~\cite{villa2024x5g} present X5G, an open, multi-vendor testbed using OpenAirInterface and NVIDIA ARC, while Upadhyaya et al.~\cite{upadhyaya2023open} develop OAIC for AI-based RAN control. Gemmi et al.~\cite{gemmi2024open6g} detail the Open6G OTIC infrastructure, which supports standardized testing of programmable RANs. These initiatives demonstrate system-level integration but often abstract away interface-specific failure analysis.

\begin{table*}[htbp]
\centering
\caption{Open RAN Testbeds and Research Platforms}
\begin{tabularx}{\textwidth}{@{}lXXX@{}}
\toprule
\textbf{Testbed Name} & \textbf{Core Components} & \textbf{Primary Focus} & \textbf{O-RAN Support (e.g., Interfaces, RIC)} \\
\midrule


\textbf{ARA}~\cite{zhang2024ara} & SDRs (e.g., Skylark), commercial base stations, DevOps tools & Rural broadband applications; Open RAN interoperability and education & E2, F1 interfaces; Near-RT RIC; Supports ONF SD-RAN \\


\textbf{Campus5G}~\cite{ferguson2025campus5g} & O-RAN-compliant private 5G, programmable RAN elements & Innovation in real-world campus environments; AI/ML integration & E2, RIC (Near-RT); Supports slicing and programmability \\

\textbf{Colosseum}~\cite{polese2024colosseum} & Large-scale emulator (128 SDRs), traffic generators, channel emulation & Massive-scale wireless experimentation; AI/ML for RAN control & E2 interface; Near-RT RIC (ColO-RAN); Integrates with OAI/srsRAN \\


\textbf{OAIC (Open AI Cellular)}~\cite{upadhyaya2023open} & OAI-based RAN, AI controllers, srsRAN & AI-based RAN management and control algorithms & E2 interface; Near-RT RIC; AI/ML xApps \\

\textbf{Open6G OTIC}~\cite{gemmi2024open6g} & Programmable RAN, multi-vendor components, testing infrastructure & Standardized testing of programmable RANs; 6G exploration & Full O-RAN (E2, O1, A1); Non-RT/Near-RT RIC; OTIC compliance \\

\textbf{OpenRAN Gym (w/ ColO-RAN)}~\cite{b11} & srsRAN/OAI, Docker-based xApps, data collection pipelines & AI/ML xApp development and deployment; End-to-end testing & E2 interface; Near-RT RIC (ColO-RAN); xApps for beamforming/load-balancing \\

\textbf{POWDER}~\cite{johnson2022nexran} & SDRs, commercial radios, cloud integration & Flexible RAN experimentation; Urban/rural scenarios & E2, RIC support; Integrates open-source stacks \\

\textbf{X5G}~\cite{villa2024x5g} & OpenAirInterface, NVIDIA ARC, multi-vendor end-to-end 5G & Programmable, multi-vendor private 5G testing & E2, F1; Near-RT RIC; Supports OAI \\
\bottomrule
\end{tabularx}
\label{tab:oran_testbeds}
\end{table*}

Yang~\cite{yang2022interoperability} introduces a testing tool for ORAN operations and maintenance interfaces, offering insight into configuration and management plane consistency. Polese et al.~\cite{polese2023understanding} provide a broad survey of ORAN architecture, algorithms, and research challenges, emphasizing the need for experimental validation. Arnaz et al.~\cite{arnaz2022toward} and Wong~\cite{wong2023open} also survey programmability and integration but stop short of in-depth analysis at the interface level.

While the experimental breadth in these studies is commendable, they are somewhat cluttered by attempts to address all O-RAN interfaces simultaneously and offer only limited systematic approaches for resolving interoperability issues. In contrast, our work narrows the scope to the E2 interface, providing an engineer-friendly perspective with practical resolution strategies and well-defined future directions for this interface.

\section{Experiment Setups}\label{sec:Experiment}

To explore practical interoperability challenges in ORAN environments, we designed and evaluated a series of experimental setups involving both commercial and open-source components.

\begin{figure}[htbp]
\centering
\includegraphics[width=\columnwidth]{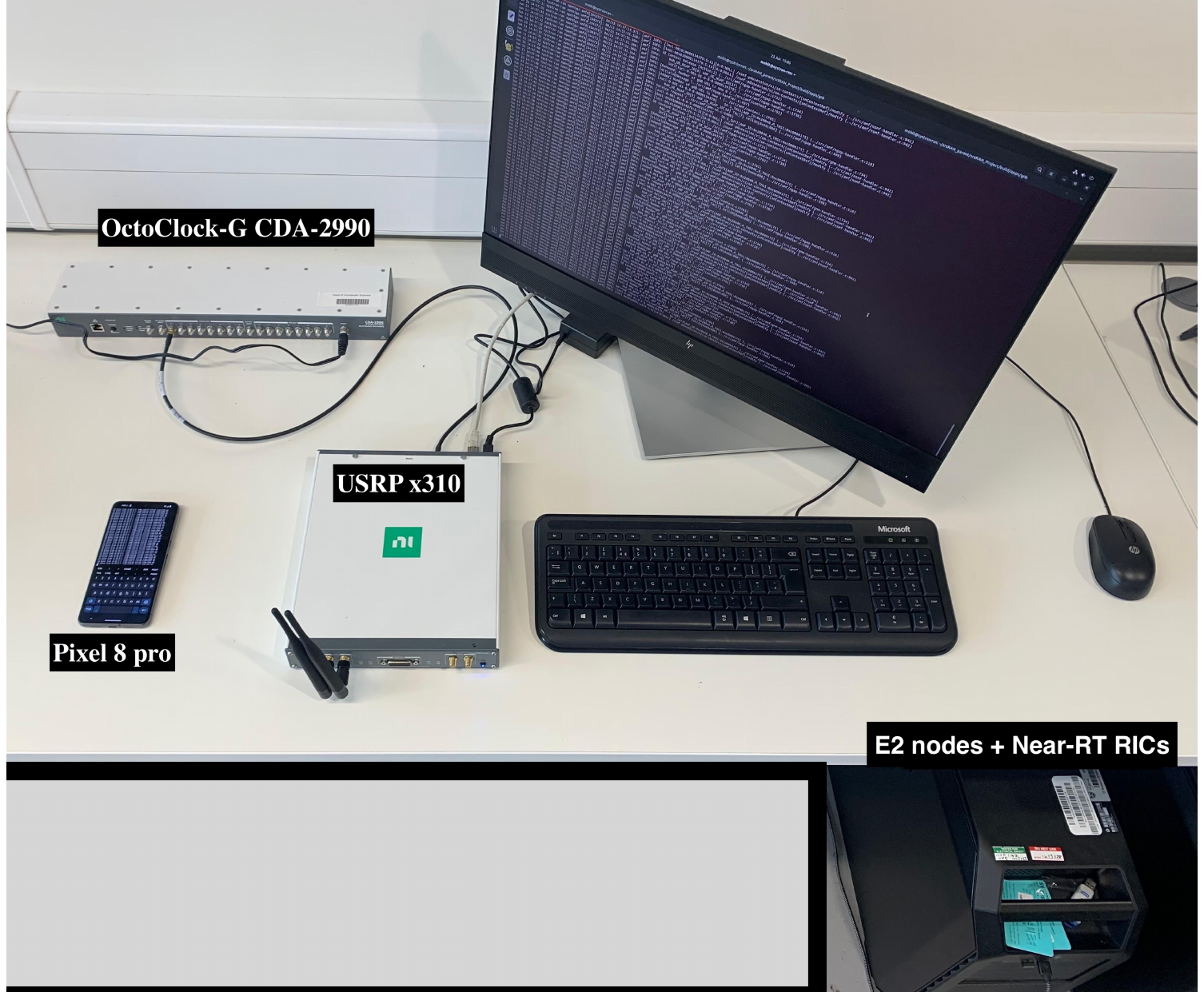}
\caption{ORAN Testbed Setup}
\label{fig:att_share_axis} 
\end{figure}

Our testbed \ref{fig:att_share_axis} focuses on the E2 interface lifecycle, particularly on how differences in service model implementation, message encoding, and capability negotiation impact integration success.

We describe four key experiment setups, each reflecting a different vendor mix and software stack. The goal is not only to validate conformance but also to uncover edge cases and protocol mismatches that hinder reliable multi-vendor deployments.

\subsection{Experiment Setup 1: Commercial E2 Node (Vendor A) with Open-Source RIC (OSC RIC)}
We utilized the OSC RIC from the O-RAN SC I-Release, compliant with E2SM-KPM v04.00, integrated with Vendor A's commercial E2 node (version 1.19, also v04.00 compliant).
\begin{figure}[H]
    \centering
    \includegraphics[width=0.8\columnwidth]{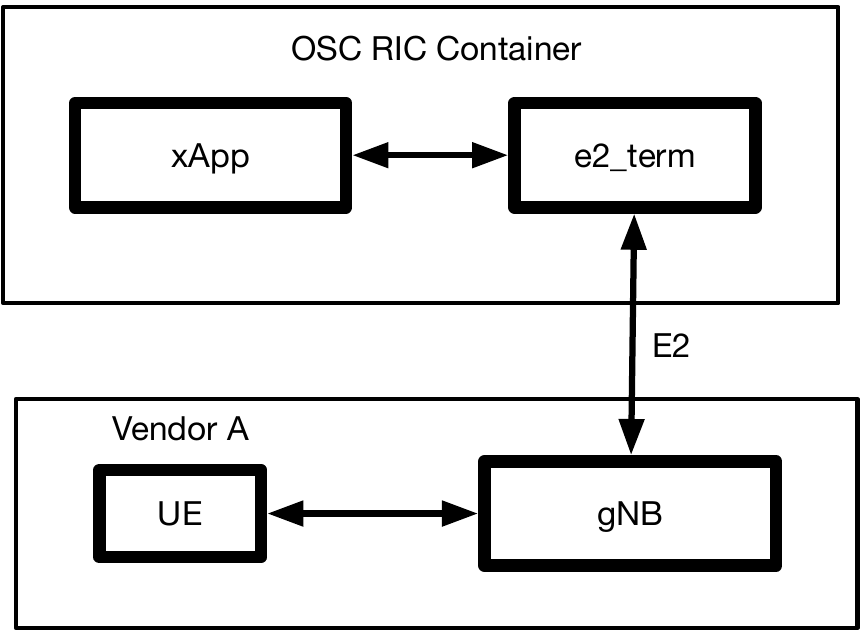}
    \caption{Setup 1 – Commercial E2 Node from Vendor A with OSC RIC}
    \label{fig:Vendor A_oscric}
\end{figure}
Initial E2 Setup procedures, including `E2 Setup Request', `E2 Setup Response', and `RIC subscription requests' were completed successfully between the OSC RIC and Vendor A's E2 node. However, the integration failed during the decoding of the `RIC Indication' message sent by the E2 node. Specifically, the OSC RIC was unable to decode the `IndicationHeader`.



Upon investigation, we identified a mismatch in the encoding expectations specified in \textit{e2sm-kpm-v4.00.asn}. The OSC RIC expected the message in IndicationHeader-Format1, whereas Vendor A’s implementation used the generic IndicationHeader. Consequently, instead of a value within the expected range [65–90, 97–122, 48–57, 32, 39–41, 43–47, 58, 61, 63], the field \textit{E2SM-KPM-IndicationHeader-Format1.vendorName} contained the value 28.
To resolve this, we modified the source code of the OSC RIC to explicitly decode the generic format as shown in Listing~\ref{lst:osc_kpm}.
\begin{lstlisting}[caption={Modification of the unpack\_indication\_header},captionpos=b, label=lst:osc_kpm]
# Original OSC RIC unpacking logic
# https://github.com/srsran/oran-sc-ric/blob/main/xApps/python/lib/asn1/e2sm_kpm_packer.py
def unpack_indication_header(self, msg_bytes):
     return self.unpack_indication_header_format1(msg_bytes)
     
def unpack_indication_header_format1(self, msg_bytes):
     indication_hdr = self.asn1_compiler.decode('E2SM-KPM-IndicationHeader-Format1', msg_bytes)
     return indication_hdr
     
#was changed to:
# Modified version to match Vendor A encoding
def unpack_indication_header(self, msg_bytes):
    indication_hdr = self.asn1_compiler.decode('E2SM-KPM-IndicationHeader', msg_bytes)
    indication_hdr = indication_hdr['indicationHeader-formats'][1]
    return indication_hdr
\end{lstlisting}
After this adjustment, the RIC successfully decoded the header and the full RIC Indication message, completing the E2 lifecycle. However, we encountered a functional limitation: only a single Key Performance Measurement (KPM) could be retrieved per subscription. This constraint results from the use of kpmReportStyle = 3, which is the only supported report style in Vendor A’s implementation. Meanwhile, the OSC RIC’s current decoding logic supports only a single KPM per subscription for this style. This highlights an important interoperability gap: xApps requiring multiple metrics are effectively blocked unless either the RIC or the node expands support for alternative report styles.

\subsection {Setup 2: Open-Source E2 Node (srsRAN) with O-RAN SC (OSC) RIC}
\begin{figure}[H]
    \centering
    \includegraphics[width=0.8\columnwidth]{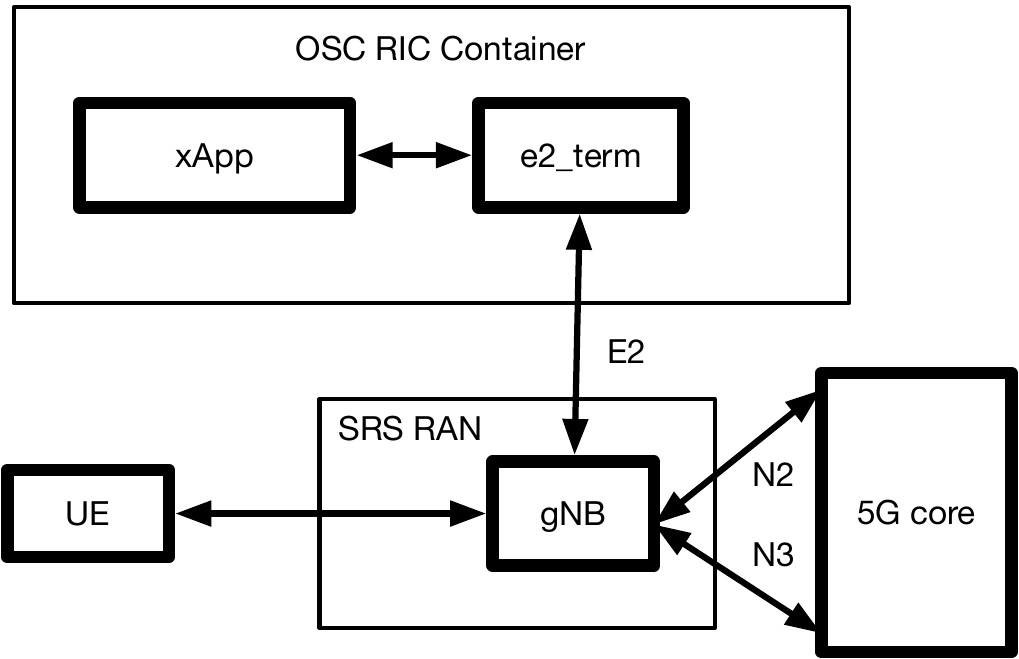}
     \caption{Setup 2 - srsRAN with OSC RIC}
    \label{fig:srsran_oscric}
\end{figure}


In contrast to the previous setup, the integration between srsRAN and the OSC RIC did not require code-level changes in our testbed. The entire E2 interface lifecycle, including setup, subscription, and data exchange, functioned as intended without modification.

Notably, srsRAN supports all kpmReportStyles (1 through 5), allowing us to subscribe to multiple KPMs within a single session, fulfilling the requirements of our xApp.

\subsection {Setup 3: Commercial E2 Node (Vendor A) with Commercial RIC (Vendor B)}
\begin{figure}[H]
    \centering
    \includegraphics[width=0.8\columnwidth]{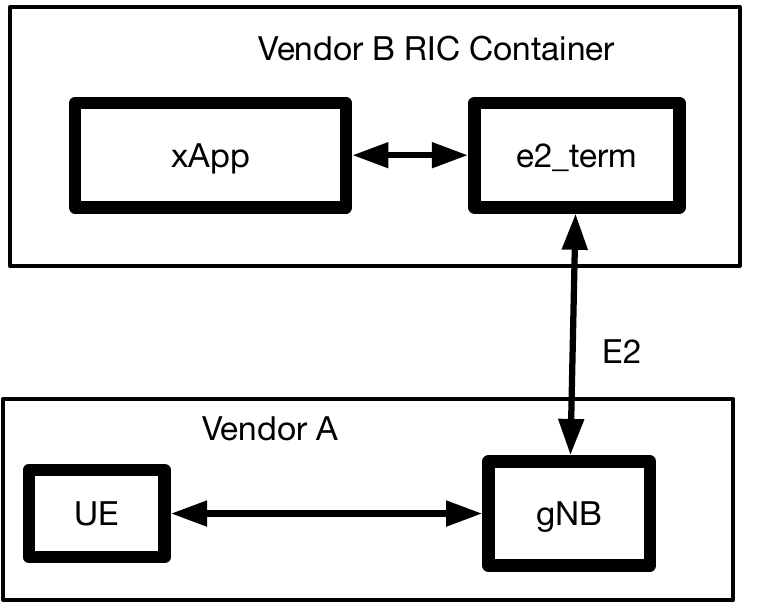}
    \caption{Setup 3: Commercial E2 Node from Vendor A with Commercial RIC from Vendor B}
    \label{fig:Vendor A_iswric}
\end{figure}
In this setup, the integration between the commercial E2 node from Vendor A and commercial RIC from Vendor B failed at the initial stage. Specifically, the E2 Setup Request/Response procedure could not be completed, preventing even the initial establishment of the E2 interface. The root cause was identified as an encoding mismatch between the two components, making them incompatible at the protocol level.

Unlike Setup 1, which used the open-source OSC RIC and allowed code-level modifications to resolve such issues, both components in this setup were commercial and closed-source. This meant no access to source code or internal configuration options to address the incompatibility.

As a result, the E2 interface could not be established, preventing any subsequent lifecycle procedures from executing. This failure, however, highlighted the strength of open-source platforms, which allow flexible customisation and rapid debugging.

\begin{table}[!htbp]
\centering
\caption{KPM Report Style Support by Vendor}
\begin{tabularx}{\linewidth}{@{}>{\raggedright\arraybackslash}p{1.8cm}>{\raggedright\arraybackslash}p{2.2cm}>{\raggedright\arraybackslash}p{4.0cm}@{}}
\toprule
\textbf{Vendor/Stack} & \textbf{Supported $kpmReportStyles$} 
& \textbf{Notes} \\
\midrule
\textbf{Vendor A (Commercial)} & Style 3 only 
& Only one KPM per subscription supported. Blocks xApps needing multiple metrics. \\

\textbf{srsRAN (Open-source)} & Style 1–5 
& Supports multiple KPMs in a single session. No source modifications needed. \\

\textbf{Vendor B RIC (Commercial)} & Unknown (E2 setup failed) 
& Integration with Vendor A failed during setup, no KPM evaluation possible. \\

\bottomrule
\end{tabularx}
\label{tab:vendor_kpm}
\end{table}

\subsection{Methodology for Root Cause Identification}
This subsection formalizes our approach as a structured,iterative framework inspired by standard software engineering practices, and the methodology consists of the following steps, applied across all setups:
\paragraph{Lifecycle Monitoring} Capture E2 interface interactions using packet analysis tools e.g. Wireshark for SCTP/ASN.1 traces and component logs e.g., RIC's outputs, node-specific telemetry.
\paragraph{Failure Isolation} Employ binary search-like segmentation of the E2 lifecycle including isolation of the setup, subscription, indication phases to pinpoint anomalies, such as message decoding errrors or dissociation events.
\paragraph{Discrepency Analysis} Compare observed behaviours against O-RAN specifications e.g., E2SM-KPM v04.00 ASN.1 schemas and vendor documentation to identify mismatches in semantics, encoding, or capabilities.
\paragraph{Root Cause Hypothesis and Testing} Formulate hypotheses e.g., format-specific expectations and test via simulations or modifications, iterating until reproducible.
\paragraph{Resolution and Validation} Implement fixes where possible and validate through re-testing, measuring metrics like resolution time, latency reduction, or success rates.
\paragraph{Documentation} Log all steps, including code snippets, traces, and metrics for reproducibility.

Table \ref{tab:open_vs_proprietary} demonstrates the comparison of root cause analysis between open-source and proprietary solutions.

\begin{table*}[!htbp]
\centering
\caption{Comparison of Root Cause Analysis Between Open-Source and Proprietary Solutions}
\begin{tabularx}
{\textwidth}{@{}lXXX@{}}
\toprule
\textbf{Aspect} & \textbf{Open-source solutions} & \textbf{Proprietary solutions} & \textbf{Impact} \\
\midrule
\textbf{Access Level} & Full source code & Limited to external logs 
& Open-source enables deeper self-directed analysis, while proprietary relies on vendor dependency, increasing resolution time (hours vs. days/weeks). \\

\textbf{Tools} & Direct code debugging, ASN.1 compilers, custom scripts.  & Vendor-provided traces, Wireshark captures, and support queries. 
& Open-source supports proactive fixes, and proprietary limited to reactive interpretation.\\

\textbf{Root Cause Depth} & Granular: Identify semantic mismatches at code level. & Surface-level: Infer causes from symptoms without internal semantics.
& Open-source yields precise insights when proprietary often inclusive without vendor input. \\

\textbf{Resolution Time/Feasibility} & Fast: 1-2 hours for fixes like encoding adjustments; 100\% resolution in tested cases. & Slow: 5+ days per query via vendor loops; some unresolved. 
& Open-source accelerates iteration while proprietary hinders scalability in research labs. \\

\textbf{Validation Approach} & Immediate re-testing with metrics.  & Dependent on vendor updates and iterative external testing.
& Open-source allows quantitative validation while proprietary qualitative and vendor-gated. \\

\bottomrule
\end{tabularx}
\label{tab:open_vs_proprietary}
\end{table*}

\section{Findings: Key Interoperability Challenges} \label{sec:Findings}
This section summarises our findings from the four ORAN setups tested in our lab.

\subsection{Vendor-Specific Implementations}

Variations in vendor-specific interpretations of the O-RAN specifications frequently lead to interoperability challenges. This was evident in Setup 1, where Vendor A and the OSC RIC used different encoding formats for the RIC Indication Header and also implemented \textit{kpmReportStyle = 3} differently. This case highlights the need for greater standardisation and clearer implementation guidelines within the O-RAN ecosystem to ensure seamless interoperability across vendors.


\subsection{Latency Sensitivity}

Real-time applications impose stringent latency requirements, demanding precise synchronization and stable communication across all components. During the integration of the Vendor A E2 node with OSC RIC, we encountered an issue where the E2 node would initially associate with the RIC but immediately dissociate, disrupting the E2AP connection lifecycle. The sequence of API calls observed during this process  is shown below:

\begin{itemize}
 \item[$\$$] {ric\_apmgr | Serving app manager}
 \item[$\$$] {ric\_rtmgr\_sim | /ric/v1/handles/e2t}
 \item[$\$$] {ric\_submgr | RMR is ready now ...}
 \item[$\$$] {ric\_rtmgr\_sim | POST /ric/v1/handles/associate-ran-to-e2t}
 \item[$\$$] {ric\_rtmgr\_sim | POST /ric/v1/handles/dissociate-ran body}
\end{itemize}


Upon investigation, the instability was attributed to latency in handling SCTP (Stream Control Transmission Protocol) `HEARTBEAT' and `HEARTBEAT ACK' messages, which are critical for maintaining stable E2AP communication. Adjustments were implemented to align the timing of these messages with the IETF RFC 4960 SCTP specification~\cite{RFC4960}. Additionally, successful ICMP PING responses towards the RIC and system clock synchronization were verified.

The issue was resolved in our setup by hosting both the Vendor A E2 node and OSC RIC on the same server, effectively mitigating latency-induced disruptions. This approach stabilized the connection and enabled the E2AP lifecycle to progress without further interruptions. However, in multi-server deployments, this latency challenge remains to be addressed.

\subsection{Multi-Node Scalability}
Scaling up to multi-node networks remains a significant challenge for open-source implementations. Synchronisation and coordination between distributed units (DUs) and centralised units (CUs) are often rudimentary, resulting in performance degradation as network size and complexity increase.

\subsection{Security Protocol Variations}
Despite ORAN’s effort to define open and interoperable interfaces, significant disparities persist in how vendors implement security protocols. These inconsistencies often arise from differing interpretations~\cite{mehran2024experimental} of security related specifications, such as authentication mechanisms, key exchange protocols, and certificate management practices. For instance, while one vendor may strictly enforce mutual TLS with X.509 certificates and periodic re-keying, another may implement a more lenient handshake process or rely on outdated cryptographic libraries. Such divergence introduces critical vulnerabilities at the interface boundaries, especially in multi-vendor deployments where secure communication is a prerequisite for reliable E2 node - RIC interaction. Without standardized enforcement or conformance testing for security protocols, these variations increase the attack surface and compromise the integrity and trust assumptions of the ORAN ecosystem.


\subsection{Limited Debugging Capabilities in Commercial Solutions}
Access restrictions on commercial RICs or E2 nodes hinder detailed debugging and customisation. For example, the failure to integrate the Vendor B RIC with Vendor A necessitated extensive collaboration with the vendor, as access to the RIC source code was unavailable. This dependency delayed issue resolution, prolonged development cycles, and reduced overall flexibility.

Open-source integration issues were resolved in approximately 2-3 hours through log inspection. The commercial integration failure remained unresolved after prolonged back and forth vendor communication due to closed-source limitations, lack of reliable up-to-date documentation and lack of visibility into internal logs or protocol handlers. 

\subsection{Standards Evolution and Implementation Lags}
The ORAN specifications are continually evolving, and vendors often adopt newer standards at varying paces. In our experiments, the srsRAN node’s broad compatibility with multiple KPM report styles demonstrated its readiness for diverse implementations, whereas Vendor A’s limited adoption of these report styles created interoperability challenges. This lag in standard adoption underscores the need for greater alignment among vendors.

\subsection{Limited Support for RAN and Cell Control}
The RICs provided extremely limited support for E2SM-RC (RAN Control) and no support for E2SM-CCC (Cell Configuration Control).

\subsection{Backward Compatibility Issues}
A significant limitation encountered was the inability to utilise configuration files from older versions of the open-source software with newer releases. This lack of backward compatibility created challenges in maintaining continuity during updates, necessitating manual reconfiguration and additional effort to adapt to the updated versions. Such limitations underscore the need for improved versioning practices and enhanced migration support to streamline software upgrades and minimise disruptions.

\section{Future Directions Worth Exploring on the road to 6G} \label{sec:Future}
The interoperability challenges identified in this work highlight fundamental limitations of the current E2 interface. Designed to enable vendor-neutral integration, E2 instead introduces new barriers due to its reliance on rigid ASN.1 encoding, static service models (E2SMs), and inconsistent vendor interpretations. In practice, these design constraints make E2 integration error-prone, difficult to extend, and sensitive to implementation variations. As demonstrated in our experiments, message decoding failures, report-style mismatches, and latency-induced instability reveal that the existing E2 framework lacks the flexibility and transparency required for large-scale, multi-vendor deployments.

To address these shortcomings, future research should explore replacing or augmenting the E2 interface with lightweight, service-based APIs built using technologies such as gRPC or REST. Unlike E2, which depends on tightly defined message structures, APIs inherently promote interoperability because they are self-describing, language-agnostic, and widely supported across software ecosystems.

An API-driven control plane could effectively eliminate many of the issues exposed in this study by:
\begin{itemize}

\item[-] Removing ASN.1 decoding dependencies and enabling flexible, self-describing data exchange through protocol buffers or JSON;

\item[-] Supporting dynamic capability discovery instead of relying on hard-coded service models;

\item[-] Enabling bi-directional streaming telemetry and low-latency control without rigid E2SM definitions;

\item[-] Simplifying debugging, versioning, and backward compatibility through mature cloud-native DevOps tools and continuous integration workflows;

\item[-] Allowing seamless integration with AI-native and intent-based management systems;

\item[-] Simplifying debugging, versioning, and backward compatibility through mature cloud-native tools;

\item[-] Allowing seamless integration with AI-native and intent-based management systems envisioned for 6G; and

\item[-] Strengthening security and trust via built-in mutual TLS, token-based authentication, and fine-grained access control at the API level.

\end{itemize}
Transitioning to an API-driven, service-based RAN architecture would preserve the logical separation between RICs and E2 nodes while providing the openness, flexibility, and automation necessary for next-generation, multi-vendor 6G deployments. This approach would not only simplify integration but also accelerate innovation across both open-source and commercial ecosystems.

\section{Acknowledgement}
This work is supported by EPSRC and DSIT funded project - CHEDDAR: Communications Hub For Empowering Distributed ClouD Computing Applications And Research (EP\-/X040518/1), (EP/Y037421/1) and EPSRC funded project REMOTE (EP/Y019229/1).
\balance
\bibliographystyle{unsrt} 
\bibliography{ref}  

@ARTICLE{b11,
  author={Polese, Michele and Bonati, Leonardo and D'Oro, Salvatore and Basagni, Stefano and Melodia, Tommaso},
  journal={IEEE Transactions on Mobile Computing}, 
  title={ColO-RAN: Developing Machine Learning-Based xApps for Open RAN Closed-Loop Control on Programmable Experimental Platforms}, 
  year={2023},
  volume={22},
  number={10},
  pages={5787-5800},
  doi={10.1109/TMC.2022.3188013}}

@inproceedings{johnson2022nexran,
  title={NexRAN: Closed-loop RAN slicing in POWDER-A top-to-bottom open-source open-RAN use case},
  author={Johnson, David and Maas, Dustin and Van Der Merwe, Jacobus},
  booktitle={Proceedings of the 15th ACM Workshop on Wireless Network Testbeds, Experimental evaluation \& CHaracterization},
  pages={17--23},
  year={2022}
}

@misc{oran_overview,
  author = {O-RAN Alliance},
  title  = {O-RAN: Towards an Open and Smart RAN},
  year   = {2020},
  note   = "[Online]. Available: https://www.o-ran.org/. Accessed: July, 2025"
}

@article{mehran2024experimental,
  title={Experimental Evaluation of Multi-Vendor 5G Open RANs: Promises, Challenges, and Lessons Learned},
  author={Mehran, Farhad and Turyagyenda, Charles and Kaleshi, Dritan},
  journal={IEEE Access},
  year={2024},
  publisher={IEEE}
}

@inproceedings{hoschele20225g,
  title={5G Interoperability of Open RAN Components in Large Testbed Ecosystem: Towards 6G Flexibility},
  author={H{\"o}schele, Thomas and Kaltenberger, Florian and Grohmann, Andreas Ingo and Tasdemir, Elif and Reisslein, Martin and Fitzek, Frank HP},
  booktitle={European Wireless 2022; 27th European Wireless Conference},
  pages={1--6},
  year={2022},
  organization={VDE}
}

@inproceedings{kord2024evolving,
  title={Evolving Open RAN Interoperability: A Large-Scale Definition},
  author={Kord, Aziz and Coder, Jason B and Le, Vu},
  booktitle={2024 IEEE International Conference on Communications Workshops (ICC Workshops)},
  pages={2095--2100},
  year={2024},
  organization={IEEE}
}

@article{bonati20245g,
  title={5G-CT: Automated Deployment and Over-the-Air Testing of End-to-End Open Radio Access Networks},
  author={Bonati, Leonardo and Polese, Michele and D'Oro, Salvatore and del Prever, Pietro Brach and Melodia, Tommaso},
  journal={IEEE Communications Magazine},
  year={2024},
  publisher={IEEE}
}

@inproceedings{zhang2024ara,
  title={ARA-O-RAN: End-to-End programmable O-RAN living lab for agriculture and rural communities},
  author={Zhang, Tianyi and Boateng, Joshua Ofori and Islam, Taimoor UI and Ahmad, Arsalan and Zhang, Hongwei and Qiao, Daji},
  booktitle={IEEE INFOCOM 2024-IEEE Conference on Computer Communications Workshops (INFOCOM WKSHPS)},
  pages={1--6},
  year={2024},
  organization={IEEE}
}

@article{polese2024colosseum,
  title={Colosseum: The open RAN digital twin},
  author={Polese, Michele and Bonati, Leonardo and D'Oro, Salvatore and Johari, Pedram and Villa, Davide and Velumani, Sakthivel and Gangula, Rajeev and Tsampazi, Maria and Robinson, Clifton Paul and Gemmi, Gabriele and others},
  journal={IEEE Open Journal of the Communications Society},
  volume={5},
  pages={5452--5466},
  year={2024},
  publisher={IEEE}
}

@article{ferguson2025campus5g,
  title={Campus5G: A Campus Scale Private 5G Open RAN Testbed},
  author={Ferguson, Andrew E and Pawar, Ujjwal and Wang, Tianxin and Marina, Mahesh K},
  journal={arXiv preprint arXiv:2506.23740},
  year={2025}
}

@article{villa2024x5g,
  title={X5G: An Open, Programmable, Multi-vendor, End-to-End, Private 5G O-RAN Testbed with NVIDIA ARC and OpenAirInterface},
  author={Villa, Davide and Khan, Imran and Kaltenberger, Florian and Hedberg, Nicholas and da Silva, Ruben Soares and Maxenti, Stefano and Bonati, Leonardo and Kelkar, Anupa and Dick, Chris and Baena, Eduardo and others},
  journal={arXiv preprint arXiv:2406.15935},
  year={2024}
}

@article{upadhyaya2023open,
  title={Open AI Cellular (OAIC): An Open Source 5G O-RAN Testbed for Design and Testing of AI-Based RAN Management Algorithms},
  author={Upadhyaya, Pratheek S and Tripathi, Nishith and Gaeddert, Joseph and Reed, Jeffrey H},
  journal={IEEE Network},
  volume={37},
  number={5},
  pages={7--15},
  year={2023},
  publisher={IEEE}
}

@inproceedings{gemmi2024open6g,
  title={Open6G OTIC: A Blueprint for Programmable O-RAN and 3GPP Testing Infrastructure},
  author={Gemmi, Gabriele and Polese, Michele and Johari, Pedram and Maxenti, Stefano and Seltser, Michael and Melodia, Tommaso},
  booktitle={2024 IEEE 100th Vehicular Technology Conference (VTC2024-Fall)},
  pages={1--5},
  year={2024},
  organization={IEEE}
}

@inproceedings{yang2022interoperability,
  title={Interoperability Testing Tool for Operations and Maintenance Interfaces of 5G Open RAN Base Station},
  author={Yang, Wan-Chien},
  booktitle={2022 23rd Asia-Pacific Network Operations and Management Symposium (APNOMS)},
  pages={1--4},
  year={2022},
  organization={IEEE}
}

@article{polese2023understanding,
  title={Understanding O-RAN: Architecture, Interfaces, Algorithms, Security, and Research Challenges},
  author={Polese, Michele and Bonati, Leonardo and D’Oro, Salvatore and Basagni, Stefano and Melodia, Tommaso},
  journal={IEEE Communications Surveys \& Tutorials},
  volume={25},
  number={2},
  pages={1376--1411},
  year={2023},
  publisher={IEEE}
}

@article{arnaz2022toward,
  title={Toward Integrating Intelligence and Programmability in Open Radio Access Networks: A Comprehensive Survey},
  author={Arnaz, Azadeh and Lipman, Justin and Abolhasan, Mehran and Hiltunen, Matti},
  journal={IEEE Access},
  volume={10},
  pages={67747--67770},
  year={2022},
  publisher={IEEE}
}

@article{wong2023open,
  title={Open RAN Test and Integration},
  author={Wong, Ian},
  journal={Open RAN: The Definitive Guide},
  pages={172--190},
  year={2023},
  publisher={Wiley Online Library}
}

@online{RFC4960,
  title        = {Stream Control Transmission Protocol},
  author       = {R. Stewart, Ed.},
  year         = {2007},
  url          = {https://datatracker.ietf.org/doc/html/rfc4960},
  note         = {Accessed: 2024-10-22},
  institution  = {IETF Network Working Group}
}
\end{document}